\documentclass{iaus}
\usepackage{graphicx}
\usepackage{amsmath}
\usepackage{natbib}

\title[Herschel far-infrared observations of M87] %% give here short title %%
{The far-infrared view of M87 \\ as seen by the Herschel
Space Observatory}

\author[Maarten Baes et al.]  %% give here short author list %%
{ M. Baes$^{1}$, M. Clemens$^{2}$, E. M. Xilouris$^{3}$,
  J. Fritz$^{1}$, W. D. Cotton$^{4}$, J. I. Davies$^{5}$,
  G. J. Bendo$^{6}$, S. Bianchi$^{7}$, L. Cortese$^{5}$,
  I. De~Looze$^{1}$, M. Pohlen$^{5}$, J. Verstappen$^{1}$,
  H. B\"ohringer$^{8}$, D. J. Bomans$^{9}$, A. Boselli$^{10}
 $, E. Corbelli$^{7}$, A. Dariush$^{5}$,
  S. di~Serego~Alighieri$^{7}$, D. Fadda$^{11}$,
  D. A. Garcia-Appadoo$^{12}$, G. Gavazzi$^{13}$,
  C. Giovanardi$^{7}$, M. Grossi$^{14}$, T. M. Hughes$^{5}$,
  L. K. Hunt$^{7}$, A. P. Jones$^{15}$, S. Madden$^{16}$,
  D. Pierini$^{8}$, S. Sabatini$^{17}$, M. W. L. Smith$^{5}$,
  C. Vlahakis$^{18}$, S. Zibetti$^{19}$
}

\affiliation{$^1$Sterrenkundig Observatorium, Universiteit Gent,
  Belgium
  \\[\affilskip]
  $^2$INAF-Osservatorio Astronomico di Padova, Italy
  \\[\affilskip]
  $^3$National Observatory of Athens, Greece
  \\[\affilskip]
  $^4$National Radio Astronomy Observatory, USA
  \\[\affilskip]
  $^5$Department of Physics and Astronomy, Cardiff University, UK
  \\[\affilskip]
  $^6$Astrophysics Group, Imperial College London, UK
  \\[\affilskip]
  $^7$INAF-Osservatorio Astrofisico di Arcetri, Firenze, Italy
  \\[\affilskip]
  $^8$Max-Planck-Institut f\"ur Extraterrestrische Physik, Garching,
  Germany
  \\[\affilskip]
  $^9$Astronomical Institute, Ruhr-University Bochum, Germany
  \\[\affilskip]
  $^{10}$Laboratoire d'Astrophysique de Marseille, France
  \\[\affilskip]
  $^{11}$NASA Herschel Science Center, California Institute of
  Technology, USA
  \\[\affilskip]
  $^{12}$European Southern Observatory, Santiago, Chile
  \\[\affilskip]
  $^{13}$Universit\`a di Milano-Bicocca, Italy
  \\[\affilskip]
  $^{14}$CAAUL, Observat\'orio Astron\'omico de Lisboa, Portugal
  \\[\affilskip]
  $^{15}$Institut d'Astrophysique Spatiale (IAS), Universit\'e Paris-Sud
  11, France
  \\[\affilskip]
  $^{16}$Laboratoire AIM, CEA/DSM, Universit\'e Paris Diderot, France
  \\[\affilskip]
  $^{17}$INAF-Istituto di Astrofisica Spaziale e Fisica Cosmica, Roma,
  Italy
  \\[\affilskip]
  $^{18}$Leiden Observatory, The Netherlands
  \\[\affilskip]
  $^{19}$Max-Planck-Institut f\"ur Astronomie, Heidelberg, Germany }

\pubyear{2010}
\volume{275}  %% insert here IAU Symposium No.
\pagerange{xxx}
\jname{Jets at all scales}
\editors{G. E. Romero, R. A. Sunyaev \& T.\ Belloni, eds.}
\begin{document}

\maketitle

\begin{abstract}
  The origin of the far-infrared emission from the nearby radio galaxy
  M87 remains a matter of debate. Some studies find evidence of a
  far-infrared excess due to thermal dust emission, whereas others
  propose that the far-infrared emission can be explained by
  synchrotron emission without the need for an additional dust
  emission component. We observed M87 with PACS and SPIRE as part of
  the Herschel Virgo Cluster Survey (HeViCS). We compare the new
  Herschel data with a synchrotron model based on infrared, submm and
  radio data to investigate the origin of the far-infrared
  emission. We find that both the integrated SED and the Herschel
  surface brightness maps are adequately explained by synchrotron
  emission. At odds with previous claims, we find no evidence of a
  diffuse dust component in M87.
 \end{abstract}

\firstsection % if your document starts with a section,

\section{Introduction}

At a distance of 16.7~Mpc%\citep{2007ApJ...655..144M}
, M87 is the dominant galaxy of the Virgo Cluster. It is one of the
nearest radio galaxies and was the first extragalactic X-ray source to
be identified. Because of its proximity, many interesting
astrophysical phenomena can be studied in more detail in M87 than in
other comparable objects%(see e.g.\ \citet{1999LNP...530.....R} for an
                        %overview)
. Particularly remarkable is the prominent jet extending from the
nucleus, visible throughout the electromagnetic spectrum. The central
regions of M87, in particular the structure of the jet, have been
studied and compared intensively at radio, optical, and X-ray
wavelengths.  \citep[e.g.][]{1991AJ....101.1632B, 1996A+A...307...61M,
  2001A+A...365L.181B, 2001ApJ...551..206P, 2004ApJ...607..294S,
  2005ApJ...627..140P, 2007ApJ...668L..27K, 2008A+A...482...97S}.

Compared to the available information at these wavelengths, our
knowledge of M87 at far-infrared (FIR) wavelengths is rather poor. A
controversial issue is the origin of the FIR emission in M87, i.e.,
the question of whether the FIR emission is caused entirely by
synchrotron emission or whether there is an additional contribution
from dust associated with either the global interstellar medium or a
nuclear dust component. %This question is partly driven by the
                        %observation of faint dust features in deep
                        %optical images \citep{1993ApJ...413..531S,
                        %2006ApJS..164..334F}. 
Several papers, based on IRAS, ISO and Spitzer observations, arrive at
different conclusions \citep{2007ApJ...663..808P, 2009ApJ...705..356B,
  2004A+A...416...41X, 2007ApJ...655..781S, 2008ApJ...689..775T}.

\begin{figure}
  \centering
  \includegraphics[width=0.56\textwidth]{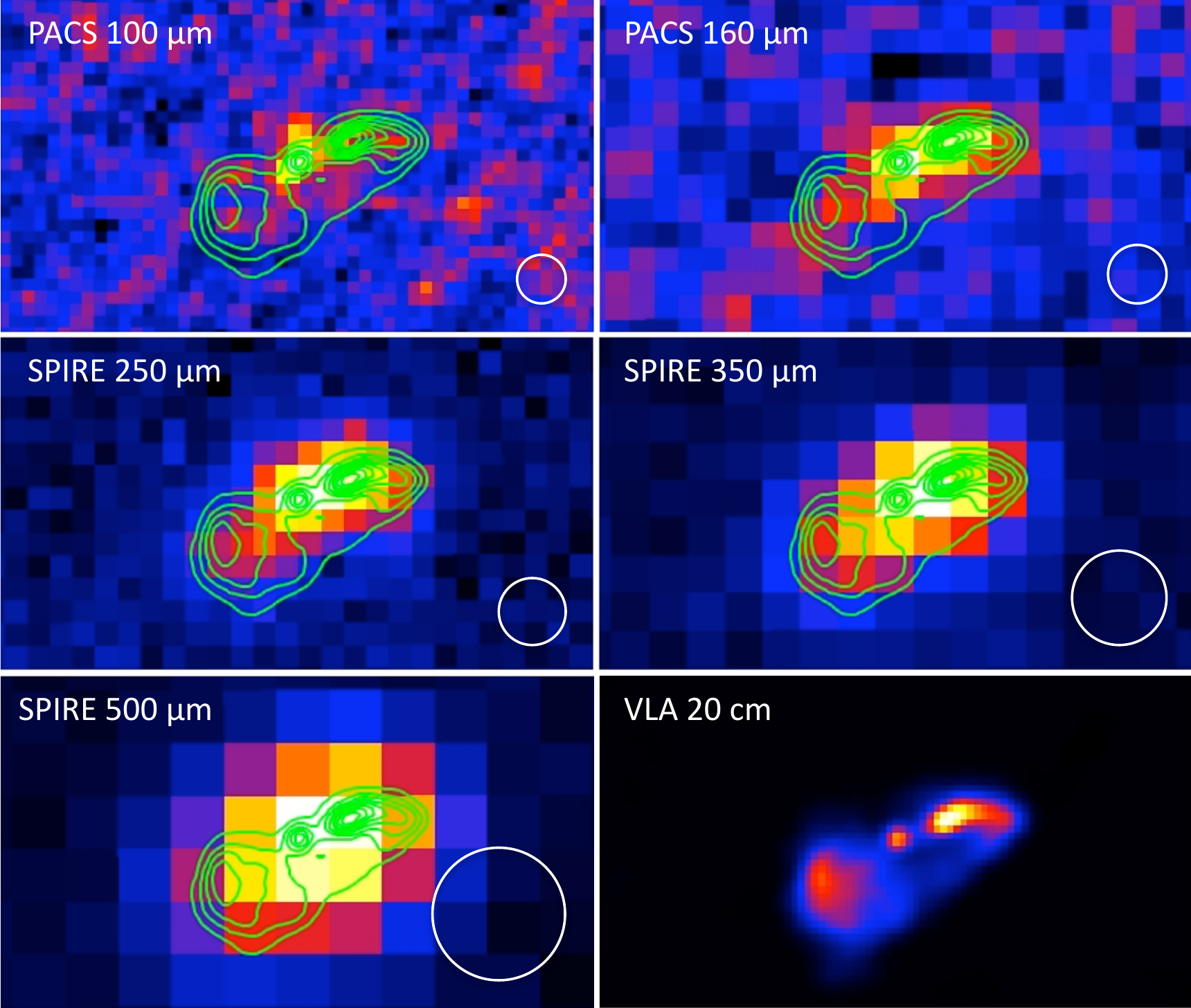}%
  \includegraphics[width=0.44\textwidth]{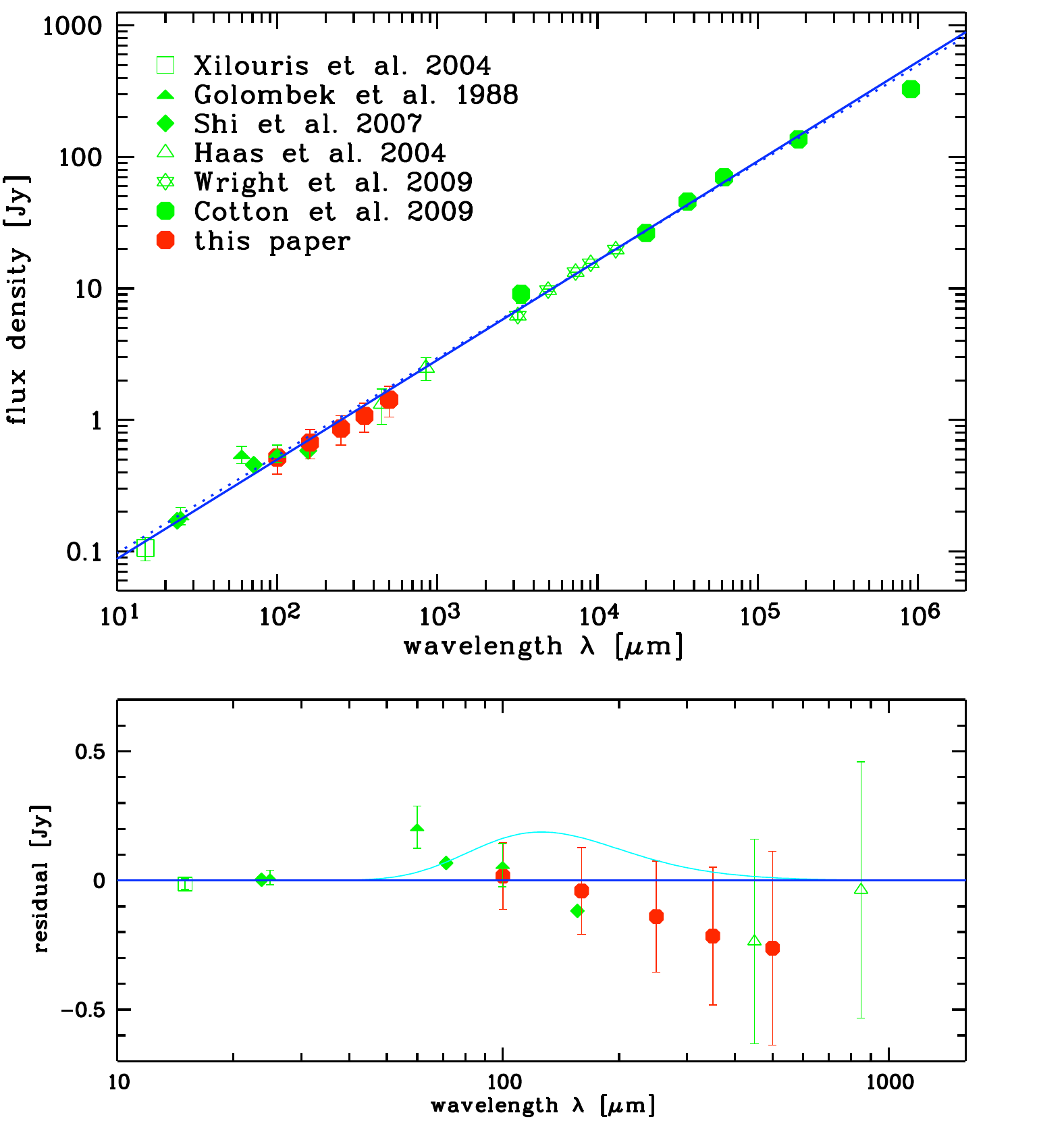}
  \caption{Left: the Herschel view of the central regions of M87. The
    bottom right image is a VLA 20~cm image from the FIRST survey. The
    20~cm radio contours have been overlaid on the Herschel
    images. The field of view of all images is $160"\times90"$, beam
    sizes are indicated in the bottom right corner. Top right: the
    global SED of M87 from mid-infrared to radio wavelengths. Where no
    error bars are seen, they are smaller than the symbol size.  The
    solid line in the plot is the best-fit power law of the ISOCAM,
    IRAS, MIPS, SCUBA, GBT, WMAP, and VLA data; the dotted line has
    only been fitted to the SCUBA, GBT, WMAP, and VLA data. Bottom
    right: residual between data and the best-fit synchrotron model in
    the infrared-submm wavelength range. The cyan line is a modified
    black-body model with $T=23$~K and
    $M_{\text{d}}=7\times10^4~M_\odot$.}
  \label{RawData.pdf}
\end{figure}
 
We investigate the nature of the FIR emission of M87 using new FIR
data from the Herschel Space Observatory \citep{2010A&A...518L...1P},
obtained as part of the science demonstration phase (SDP) observations
of the Herschel Virgo Cluster Survey
\citep[HeViCS,][]{2010A&A...518L..48D}. HeViCS is an approved open
time key program, which has been awarded 286~h of observing time in
parallel mode with the PACS and SPIRE instruments. We will ultimately
map four $4\times4$ square degree regions of the cluster at 100, 160,
250, 350 and 500~$\mu$m, down to the 250~$\mu$m confusion limit of
about 1~MJy~sr$^{-1}$. The HeViCS SDP observations consisted of a
single cross-scan of one $4\times4$~deg$^2$ field at the centre of the
Virgo Cluster. While these SDP observations comprise only 6\% of the
total HeViCS observations, the analysis based on these observations
already gives a prelude to the primary HeViCS science goals including
the effects of the environment on the dust medium of galaxies, the FIR
luminosity function, the complete SEDs of galaxies and a detailed
analysis of the dust content of dwarf and early-type galaxies
\citep{2010A&A...518L..48D, 2010A&A...518L..49C, 2010A&A...518L..50C,
  2010A&A...518L..51S, 2010A&A...518L..52G, 2010A&A...518L..54D,
  2010A&A...518L..61B}. The observations also enabled us to study the
intensity and nature of the FIR emission of M87
\citep{2010A&A...518L..53B}.

\section{Analysis and conclusion}
\label{Analysis.sec}

The left part of Figure~{\ref{RawData.pdf}} shows the PACS and SPIRE
images of the central $160"\times90"$ region of M87, which is clearly
detected in all five bands. The top right panel in
Figure~{\ref{RawData.pdf}} shows the integrated SED in the
infrared-submm-radio region between 15~$\mu$m and 100~cm, with the new
Herschel data as well as ISOCAM, IRAS, MIPS, and SCUBA, GBT, WMAP, and
VLA data gathered from the literature \citep{2004A+A...416...41X,
  1988AJ.....95...26G, 2007ApJ...655..781S, 2004A+A...424..531H,
  2009ApJ...701.1872C, 2009ApJS..180..283W}; the actual flux densities
can be found in \citet{2010A&A...518L..53B}. The solid line is the
best-fit power law for all literature data and has a slope
$\alpha=-0.76$; the dotted line fits only the submm and radio data and
has a slope $\alpha=-0.74$. The bottom right panel in
Figure~{\ref{RawData.pdf}} shows the residual from the best-fit power
law in the infrared-submm wavelength region; clearly, the integrated
Herschel fluxes are in full agreement with synchrotron radiation. The
cyan line in this figure is a modified black-body fitted with $T=23$~K
and $M_{\text{d}}=7\times10^4~M_\odot$. This temperature is the mean
dust equilibrium temperature in the interstellar radiation field of
M87, determined using the SKIRT radiative transfer code
\citep{2003MNRAS.343.1081B, 2005AIPC..761...27B} and based on the
photometry from \citet{2009ApJS..182..216K}. The dust mass was
adjusted to fit the upper limits of the residuals. It is clear that
the SED of M87 is incompatible with dust masses higher than
$10^5~M_\odot$.

\begin{figure}
  \centering
  \includegraphics[width=\textwidth]{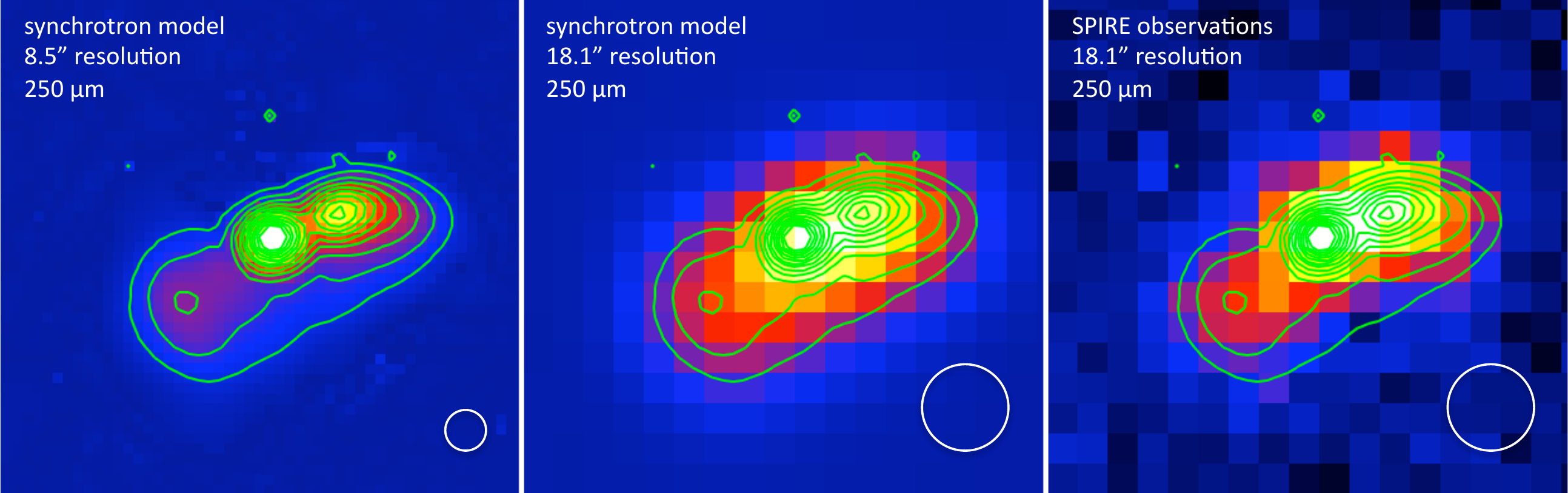}
  \caption{A comparison between the synchrotron model image and the
    observed image at 250~$\mu$m. The left panel shows the synchrotron
    image at the model resolution, the central panel shows the same
    model convolved to the SPIRE 250~$\mu$m beam and pixel size. The
    right panel shows the observed SPIRE 250~$\mu$m image. In all
    panels, the green lines are the contours of the synchrotron model
    at the model resolution.}
  \label{H250.pdf}
\end{figure}
 
Although indicative, the analysis of the integrated SED does not
definitively identify the origin of the FIR emission in
M87. Approximating the global SED as a single power-law synchrotron
model is indeed an oversimplification of the complicated structure of
M87. Several studies have shown that M87 contains three distinct
regions of significant synchrotron emission, each with their own
spectral indices: the nucleus, the jet and associated lobes in the NW
region, and the SE lobes \citep[e.g.,][]{1991AJ....101.1632B,
  1996A+A...307...61M, 2001ApJ...551..206P, 2007ApJ...655..781S}. We
have constructed a synchrotron model for the central regions of M87,
based on a newly reduced MIPS 24~$\mu$m map and archival MUSTANG
90~GHz and 15, 8.2, 4.9, 1.6, and 0.3~GHz maps
\citep{2009ApJ...701.1872C}. We fitted a second-order polynomial
synchrotron model to each pixel of the MIPS + radio data cube and used
this synchrotron model to predict the emission of M87 at 250~$\mu$m
(the SPIRE 250~$\mu$m image provides the optimal compromise between
S/N and spatial resolution). Figure~{\ref{H250.pdf}} shows the
comparison between the synchrotron model prediction at 250~$\mu$m and
the SPIRE observations. At the model resolution, the three distinct
components are visible, but when we convolve this synchrotron model
image with the SPIRE 250~$\mu$m beam, the three different components
merge into a single extended structure with one elongated peak
slightly west of the nucleus. Comparing the central and right panels
of Figure~{\ref{H250.pdf}}, we see that the synchrotron model is
capable of explaining the observed SPIRE 250~$\mu$m image
satisfactorily.

We conclude that for both the integrated SED and the SPIRE 250~$\mu$m
map, we have found that synchrotron emission is an adequate
explanation of the FIR emission. We do not detect a FIR excess that
cannot be explained by the synchrotron model. In particular, we have
no reason to invoke the presence of smooth dust emission associated
with the galaxy interstellar medium, as advocated by
\citet{2007ApJ...655..781S}. For a dust temperature of 23~K, which is
the expected equilibrium temperature in the interstellar radiation
field of M87, we find an upper limit to the dust mass of
$7\times10^4~M_\odot$. % Our result
% agrees with the analysis of the nuclear emission by
% \citet{2009ApJ...705..356B}. \citet{Clemens} discuss the lifetimes
% of interstellar dust grains in elliptical galaxies in the Virgo
% Cluster based on Herschel observations and find an upper limit to
% the amorphous silicate grain survival time of less than 46 million
% years. Given that M87 is a luminous X-ray source, the absence of a
% substantial dust component is not a surprise. A low dust content is
% also in agreement with the non-detection of cool molecular gas
% \citep{2008A+A...489..101S, 2008ApJ...689..775T} and the
% non-detection of significant intrinsic absorption in the X-ray
% spectra of M87 \citep{2002A+A...382..804B}.
Our conclusion is that, seen from the FIR point of view, M87 is a
passive object with a central radio source emitting synchrotron
emission, without a substantial diffuse dust component.

% \begin{acknowledgement}
%   The National Radio Astronomy Observatory (NRAO) is operated by
%   Associated Universities Inc, under cooperative agreement with the
%   National Science Foundation.
% \end{acknowledgement}

\end{document}